\documentclass[12pt]{article}

\usepackage{cite}
\usepackage{axodraw}
\usepackage[dvips]{graphicx}
\usepackage{epsfig}

\def\g{{\rm I}\hspace{-0.07cm}\Gamma}

\begin{document}

\begin{flushright}
MAN/HEP/2008/23\\[-2pt]
{\tt arXiv:0809.1580}\\
September 2008
\end{flushright}
\bigskip

\begin{center}
{\Large {\bf The 1-Loop Effective Potential in Non-Linear Gauges}}\\[1.5cm] 
{\large Lisa P. Alexander and Apostolos Pilaftsis }\\[0.5cm]
{\em School of Physics and Astronomy, University of Manchester,}\\ 
{\em Manchester M13 9PL, United Kingdom }
\end{center}

\vspace{1.5cm}  
\centerline{\bf  ABSTRACT}  

\noindent
We calculate the 1-loop effective  potential of an Abelian Higgs model
within the $R_{\xi/\sigma}$ class  of non-linear gauges that preserves
the  Higgs-boson  low-energy   theorem.   The  $R_{\xi/\sigma}$  gauge
involves  two gauge-fixing  parameters $\xi$  and $\sigma$,  and  is a
renormalizable  extension  of the  Feynman--'t  Hooft  $R_\xi$ set  of
gauges beyond  the 1-loop level.  By taking  consistently into account
Goldstone--gauge-boson  mixing   effects,  we  show   how  the  1-loop
effective potential  evaluated at its  extrema is independent  of both
$\xi$  and $\sigma$,  in  agreement with  the  Nielsen identity.   The
1-loop constant and wavefunction renormalizations are presented within
this general $R_{\xi/\sigma}$ class of non-linear gauges.

\medskip
\noindent
%{\small PACS numbers: }

\newpage

\setcounter{equation}{0}
\section{Introduction}

In quantum  field theory,  the effective potential~\cite{CW}  plays an
important  role.  It  carries  precious information  about the  vacuum
structure  of   a  given  model~\cite{GSW,J-L},   and  determines  its
stability  under   quantum  fluctuations.   The   effective  potential
constitutes a useful field-theoretic tool for studying the dynamics of
inflationary  cosmology~\cite{inflation}  and  for  investigating  the
order  of  the  phase  transition  at  finite  temperatures~\cite{FT}.
Conventionally,   the  effective   potential  is   denoted  as~$V_{\rm
eff}(\phi)$ and  is a functional  of all classical fields  involved in
the theory that are collectively represented here by $\phi$.

An important property of  the effective potential~$V_{\rm eff} (\phi)$
is its  independence on  the gauge-fixing parameter,  e.g.~$\xi$, when
evaluated  for classical  field values  $\phi$ that  extremize $V_{\rm
eff} (\phi)$.   In particular, the minimum of  the effective potential
$V_{\rm  eff} (\phi_{\rm  min})$ corresponds  to a  physical quantity,
that  is the  energy  density  of the  vacuum.   Therefore, the  gauge
independence of $V_{\rm eff} (\phi_{\rm min})$ is only a manifestation
of  this  fact.   Formally,   the  gauge  dependence  of  one-particle
irreducible (1PI) $n$-point correlation functions $\g_{\phi^n}$, which
involve  a number  $n$ of  external $\phi$  fields, is  governed  by a
functional  differential  equation,  the  so-called  Nielsen  identity
(NI)~\cite{Nielsen}.  The NI relates  the explicit gauge dependence of
the  effective action  $\g[\phi,  \xi]$  to a  non-linear  sum of  the
functional derivatives  $\delta \g/\delta \phi$ for  all fields $\phi$
that carry non-zero charges under  the gauge group.  Evidently, at the
extremal points where $\delta \g/\delta  \phi = 0$, both the effective
action and the resulting effective potential are gauge invariant.

In a gauge theory with spontaneous symmetry breaking (SSB), as the one
that  we   will  be   considering  here,  one   has  to   specify  the
gauge-fixing~(GF) scheme in order  to eliminate the unphysical degrees
of freedom from the gauge  fields.  One of the frequently discussed GF
schemes  in the  literature is  the Feynman--'t  Hooft $R_\xi$  set of
linear gauges, which may be described by the Lagrangian term
\begin{equation}
  \label{Rxi}
{\cal L}_{\rm GF}^{R_{\xi}}\ =\ 
\frac{1}{2\xi} \bigg(\partial_{\mu} A^{\mu}\: +\: \frac{1}{2}\;\xi g v
G\,\bigg)^{2}\; ,
\end{equation}
where $v$  is the  vacuum expectation value  (VEV) of a  complex field
$\Phi$.   The GF  choice~(\ref{Rxi}) has  the virtue  that  the mixing
between the gauge field~$A^\mu$  and its associated would-be Goldstone
boson $G$ is absent  at the tree-level, thereby simplifying enormously
calculations.   However, as we  will see  here, quantum  effects spoil
this nice property beyond the 1-loop level.

Another  disadvantage of  the $R_\xi$  set of  gauges is  that 1-loop
effects  mediated by  gauge interactions~\cite{Pilaftsis}  violate the
so-called           Higgs-boson           low-energy           theorem
(HLET)~\cite{HLET,HLETreview}, as expressed by the simple relation:
\begin{equation}
  \label{HLET}
\g_{H^{n+1}}\ =\ \frac{\partial \g_{H^{n}}}{\partial v}\; ,
\end{equation}
where $H$  is the radial  quantum excitation of $\Phi$,  more commonly
known as the Higgs boson.  Note that the additional Higgs field $H$ in
$\g_{H^{n+1}}$  occuring   on  the  LHS   of~(\ref{HLET})  carries  no
momentum.  The violation of  the HLET~(\ref{HLET}) is a consequence of
an  explicitly breaking  of a  dilatational or  translational symmetry
that governs the gauge-invariant part of the Lagrangian. Specifically,
in  the absence of  ${\cal L}_{\rm  GF}^{R_{\xi}}$, the  Lagrangian is
invariant under the translational transformation~\cite{Pilaftsis}
\begin{equation}
  \label{trWI}
H \to H + a, \qquad v \to v-a\; ,
\end{equation}
where   $a$   is   an   arbitrary   real  constant.    As   has   been
shown~\cite{Binosi}, the  translational symmetry~(\ref{trWI}) not only
implies that  the HLET~(\ref{HLET}) is manifestly  satisfied, but also
that  the  VEV $v$  gets  only  multiplicatively  renormalized by  the
wavefunction of the Higgs field to all orders in perturbation theory.

In this  paper we study the  gauge dependence of  the 1-loop effective
potential of  an Abelian  Higgs model in  a class  $R_{\xi/\sigma}$ of
non-linear  gauges.   More  explicitly,   the  GF  Lagrangian  of  the
$R_{\xi/\sigma}$ gauges~\cite{Kastening,Pilaftsis,Binosi} is given by
\begin{equation}
  \label{Rxisigma}
\mathcal{L}_{\rm GF}\ =\ -\; 
\frac{\sigma}{2\xi}\; \Big[\, \partial_{\mu}A^{\mu}\: +\: 
\xi g (H+v)G\, \Big]^2\; .
\end{equation}
The $R_{\xi/\sigma}$  set of gauges contains two  GF parameters: $\xi$
and $\sigma$. For $\sigma =1$, this is a non-linear realization of the
Feynman--'t  Hooft  gauge  that  preserves  the  HLET\footnote{Earlier
studies  have used  linear  gauges to  study  the effective  potential
~\cite{Metaxas}.    However,   these  gauges   do   not  respect   the
HLET~(\ref{HLET}) in  the $R_\xi$-gauge limit  given by~(\ref{Rxi}).}.
However, quantum  corrections spoil the condition $\sigma  = 1$ beyond
the 1-loop level,  and a new GF parameter  is required to consistently
provide a renormalizable extension of the $R_\xi$ gauges.  To the best
of  our knowledge,  there is  no  explicit calculation  of the  1-loop
effective potential in the  $R_{\xi/\sigma}$ set of non-linear gauges,
for arbitrary  $\xi$ and $\sigma$  parameters.  Such a  calculation is
not trivial, since one has to properly take into account the mixing of
the gauge field with the respective would-be Goldstone boson.  We then
show that the 1-loop effective potential at its extrema is independent
of  the GF  parameters  $\xi$ and  $\sigma$,  as required  by the  NI.
Finally,   we  present  the   1-loop  renormalization   constants  for
couplings,  masses and  wavefunctions of  all fields  involved  in the
theory.
 
The  paper  is  structured  as  follows:  Section~\ref{model}  briefly
reviews  the  Abelian Higgs  model  and  its  quantization within  the
$R_{\xi/\sigma}$ class  of non-linear gauges. In the  same section, we
present  the relevant  propagator  matrix that  takes  account of  the
mixing between the  gauge field and the would-be  Goldstone boson.  In
Section~\ref{EP} we calculate the  1-loop effective potential for both
$\sigma =1$ and  $\sigma \neq 1$.  In Section~\ref{GFP}  we extend the
derivation of the NI for the two GF parameters $\xi$ and $\sigma$, and
show how the 1-loop effective  potential at its extrema is independent
of these  two parameters.  In Section~\ref{1LR} we  present the 1-loop
counterterms  for the field  wavefunctions and  for the  couplings and
masses  for   arbitrary  values  of  $\xi$   and  $\sigma$.   Finally,
Section~\ref{conclusions} contains our conclusions.

\setcounter{equation}{0}
\section{The Abelian Higgs Model in the {\boldmath $R_{\xi / \sigma}$}
  Gauge}\label{model}

Here  we  briefly  review  the  Abelian Higgs  model  along  with  its
quantization in the  $R_{\xi / \sigma}$ gauge~\cite{Kastening,Binosi}.
In detail,  the Lagrangian  for the  Abelian Higgs model  is a  sum of
three terms,
\begin{equation}\label{eqn:L}
  \label{Lagr}
\mathcal{L}\ =\ \mathcal{L}_{\rm inv}\: +\: \mathcal{L}_{\rm GF}\: +\:
\mathcal{L}_{\rm FP}\; ,
\end{equation}
where $\mathcal{L}_{\rm inv}$  is the $U(1)_{Y}$ gauge-invariant part,
${\cal L}_{\rm  GF}$ is the GF  Lagrangian and ${\cal  L}_{\rm FP}$ is
the  induced  Faddeev--Popov  (FP)  Lagrangian  describing  the  ghost
interactions.  The gauge-invariant part of the Lagrangian reads:
\begin{eqnarray}\label{eqn:L_{I}}
\mathcal{L}_{\rm inv} \!& = &\! -\frac{1}{4}F^{\mu\nu}F_{\mu\nu} +
 (D_{\mu}\Phi)^{*}(D^{\mu}\Phi) +
 \sum_{i=1,2}[\bar{f}_{iL}(i\gamma^{\mu}D_{\mu})f_{iL} +
 \bar{f}_{iR}(i\gamma^{\mu}\partial_{\mu})f_{iR}] -
 m^{2}\Phi^{*}\Phi \nonumber\\ & & - \lambda(\Phi^{*}\Phi)^{2} -
 \sqrt{2}h_{1}(\bar{f}_{1L}\Phi f_{1R}+ \bar{f}_{1R}\Phi^{*}
 f_{1L}) - \sqrt{2}h_{2}(\bar{f}_{2L}\Phi^{*} f_{2R} +
 \bar{f}_{2R}\Phi f_{2L})\; ,
\end{eqnarray}
where $F_{\mu \nu} =  \partial_{\mu} A_{\nu} - \partial_{\nu} A_{\mu}$
is  the  U(1)$_Y$  field-strength  tensor, $D_{\mu}  =  \partial_{\mu}
-igYA_{\mu}$    is   the    covariant   derivative    and    $\Phi   =
\frac{1}{\sqrt{2}}(v + H + iG)$  is a complex field.  As was mentioned
in the introduction, $v$ is the  VEV of $\Phi$, $H$ is the Higgs field
and $G$  is the would-be  Goldstone boson of $A^\mu$.   The U(1)$_{Y}$
quantum numbers for the various fields are $Y_{\Phi}=1$, $Y_{f_{1L}} =
-Y_{f_{2L}}  = 1$ and  $Y_{f_{1R}} =  Y_{f_{2R}} =  0$. Note  that the
left-handed chiral  fermions fall into  a non-anomalous representation
by having opposite U(1)$_{Y}$ charges.

The GF term in~(\ref{Lagr}) is given by
\begin{equation}
  \label{eqn:L_{GF}}
\mathcal{L}_{\rm GF}\ =\ \frac{\xi}{2 \sigma}B^{2}\: +\: BF\; ,
\end{equation}
where $F  = \partial_{\mu}A^{\mu}  +\xi g (H+v)G$  is a  non-linear GF
function pertinent to the $R_{\xi/\sigma}$  class of gauges and $B$ is
an  auxiliary  field  which  is  introduced  to  close  the  so-called
Becchi--Rouet--Stora (BRS) transformations~\cite{BRS} off-shell.  Upon
integrating  out the  $B$ fields,  we  obtain the  GF Lagrangian  given
in~(\ref{Rxisigma}).

Finally, the FP ghost term is induced by the GF function $F$ as follows:
\begin{equation}
  \label{eqn:L_{FP}}
\mathcal{L}_{\rm FP}\ =\ -\bar{c}(sF)
\end{equation}
where  $c~(\bar{c})$ is the  ghost~(anti-ghost) field  and $s$  is the
anticommuting BRS operator~\cite{BRS}. The action of $s$ on the fields
is given by
\begin{eqnarray}
  \label{eqn:BRStransformations}
&& sA_{\mu}\ =\ \partial_{\mu}c, \quad  sH\ =\ -gcG, \quad  sG\ =\ gc(H+v),
  \quad sB\ =\ 0, \quad sc\ =\ 0,   \nonumber\\ 
&& s\bar{c}\ =\ B, \quad sf_{L}\ =\ igY_{f_L}\,cf_{L},  \quad
  s\bar{f}_{L}\ =\ igY_{f_{L}}\bar{f}_{L}c, \quad sf_{R}\ =\ 0, \quad
  s\bar{f}_{R}\ =\ 0\; . \qquad
\end{eqnarray} 
Given these BRS transformations, the FP ghost term becomes
\begin{equation}
\mathcal{L}_{\rm FP}\ =\ 
-\bar{c} \{ \partial^{2} + g^{2}\xi [(H+v)^{2} -G^{2}] \} c\; .
\end{equation}

In  the $R_{\xi  / \sigma}$  gauge, the  full bare  Lagrangian  of the
Abelian  Higgs  model  may  be  represented  as  a  sum  of  5  terms,
i.e.~${\cal L}  = \sum_{n=0}^5\; {\cal L}_n$, where  the subscript $n$
denotes the  number of the quantum fields  involved.  More explicitly,
the individual terms of the sum are given by
\begin{eqnarray}
  \label{eqn:L_0}
\mathcal{L}_{0} \!& = &\! 
-\frac{v^{2}}{2}(m^{2} + \lambda \frac{v^{2}}{2})\; ,\\
  \label{eqn:L_{1}} 
\mathcal{L}_{1} \!& = &\! - vH(m^{2} + \lambda v^{2})\; ,\\
  \label{eqn:L_2}
\mathcal{L}_{2} \!& = &\!  
\frac{1}{2}\partial_{\mu}H\partial^{\mu} H - \frac{1}{2}(3\lambda
v^{2}+m^{2})H^{2}+ \frac{1}{2}\partial_{\mu}G\partial^{\mu}G -
\frac{1}{2}[(\lambda +\sigma \xi g^{2})v^{2} + m^{2}]G^{2} \nonumber\\ 
 & & -\frac{1}{4}F_{\mu\nu}F^{\mu\nu} -
\frac{\sigma}{2\xi}(\partial_{\mu}A^{\mu})^{2} +
\frac{1}{2}g^{2}v^{2}A_{\mu}A^{\mu} -
(\sigma-1)gvG(\partial_{\mu}A^{\mu}) +
\partial_{\mu}\bar{c}\partial^{\mu}c \nonumber\\ 
 & & - \xi g^{2}v^{2}\bar{c}c +
\sum_{i=1}^2\bar{f}_{i}(i\gamma_{\mu}\partial^{\mu} - h_{i}v)f_{i}\; ,\\ 
  \label{eqn:L_3}
\mathcal{L}_{3} \!& = &\! (\sigma+1)gA_{\mu}G\partial^{\mu}H +
 (\sigma-1)gA_{\mu}H\partial^{\mu}G + g^{2}vA_{\mu}A^{\mu}H -\lambda
 vH^{3}  - 2\xi g^{2}vH\bar{c}c \nonumber\\ 
 & & - (\lambda + \sigma \xi g^{2})vHG^{2} +
 \sum_{i=1}^2\bar{f}_{i}(gA^{\mu}Y^{L}_{i}\gamma_{\mu}P_{L} - h_{i}H -
 Y^{L}_{i}h_{i}\gamma_{5}G)f_{i}\; ,\\   
  \label{eqn:L_4}
\mathcal{L}_{4} \!& = &\! \frac{1}{2}g^{2}A_{\mu}A^{\mu}H^{2} +
 \frac{1}{2}g^{2}A_{\mu}A^{\mu}G^{2} - \frac{\lambda}{4}H^{4} -
 \frac{1}{2}(\lambda + \sigma \xi g^{2} ) H^{2}G^{2} -
 \frac{\lambda}{4}G^{4} \nonumber\\ 
 & & - \xi g^{2} H^{2}\bar{c}c + \xi g^{2} G^{2}\bar{c}c\; .  
\end{eqnarray}

The  propagators of  the theory  may be  calculated from  ${\cal L}_2$
in~(\ref{eqn:L_2}).  We first consider the gauge sector, whose kinetic
Lagrangian may be cast into the matrix form:
\begin{equation} 
  \label{eqn:matrix}
\mathcal{L}_{2}^{G,A_{\mu}}\ =\ \frac{1}{2}(G, A^{\alpha})
\left( \begin{array}{cc}
-\partial_{\mu}\partial^{\mu}- m_{G}^{2} & -(\sigma-1)gv\partial_{\beta}\\
(\sigma-1)gv\partial_{\alpha} &
(\partial^{2}+m_{A}^{2})g_{\alpha\beta} +
\bigg(\frac{\displaystyle\sigma}{\displaystyle \xi}-1\bigg)\,
\partial_{\alpha}\partial_{\beta} 
\end{array}\right)
\left( \begin{array}{c} G \\ A^{\beta} \end{array}\right)\, .
\end{equation}
Notice that for $\sigma \neq 1$, there is a non-trivial mixing between
the  gauge  field  $A^\mu$  and  the  would-be  Goldstone  boson  $G$.
Inverting the matrix  in~(\ref{eqn:matrix}), one finds the propagators
in the momentum space:
\begin{eqnarray} 
  \label{eqn:AGpropsigmanotone}
D_{GG}(k) \!&=&\!  
\frac{k^{2}-
  (m_{c}^{2}/\sigma)}{(k^{2} -
  \bar{m}_{G}^{2})(k^{2}-\bar{m}_{c}^{2})}\ , \nonumber\\  
D_{A^{\mu} A^{\nu}}(k) \!&=&\! 
\frac{1}{k^{2} - m_{A}^{2}}\left(- g^{\mu\nu} +
\frac{k^{\mu}k^{\nu}}{k^{2}} \right)\ -\ \frac{\xi}{\sigma}\
\frac{k^{2} - m_{G}^{2}}{(k^{2}-\bar{m}_{G}^{2})(k^{2}-\bar{m}_{c}^{2})}\ 
\frac{k^{\mu}
  k^{\nu}}{k^2}\ ,\\ 
D_{GA^{\mu}}(k) \!&=&\!
\frac{\xi}{\sigma}\; 
\frac{igv(\sigma-1)\,k^{\mu}}{(k^{2}
  -\bar{m}_{G}^{2})(k^{2}-\bar{m}_{c}^{2})}\ , 
\qquad D_{A^{\mu}G}(k)\  =\
-\;\frac{\xi}{\sigma}\; \frac{igv (\sigma-1)\,k^{\mu}}{(k^{2}
  -\bar{m}_{G}^{2})(k^{2}-\bar{m}_{c}^{2})}\ .\nonumber\qquad   
\end{eqnarray}
Observe  that  there  is  a  relative minus  sign  between  the  mixed
propagators  $D_{GA^{\mu}}(k)$   and  $D_{A^{\mu}G}(k)$  due   to  the
different direction  of momentum flow  signified by the  subscripts $G
A^\mu$  and  $A^\mu G$.   In  (\ref{eqn:AGpropsigmanotone}), $m^2_A  =
g^2v^2$ is  the physical  gauge-boson mass, and  $\bar{m}_{G}^{2}$ and
$\bar{m}_{c}^{2}$ are  unphysical parameters related  to the Goldstone
and ghost masses:
\begin{eqnarray}
  \label{eqn:2.16}
\bar{m}_{G}^{2} \!& = &\! \frac{1}{2}\;[m_{G}^{2} + m_{c}^{2}(2 - \sigma)]\ +\
\frac{1}{2}\;\sqrt{[m_{G}^{2} + m_{c}^{2}(2 - \sigma)]^{2} -
  4\frac{m_{G}^{2}m_{c}^{2}}{\sigma}}\ ,   \nonumber\\ 
\bar{m}_{c}^{2} \!& = &\! \frac{1}{2}\;[m_{G}^{2} + m_{c}^{2}(2 - \sigma)]\ -\
\frac{1}{2}\;\sqrt{[m_{G}^{2} + m_{c}^{2}(2 - \sigma)]^{2} -
  4\frac{m_{G}^{2}m_{c}^{2}}{\sigma}}\ ,
\end{eqnarray}
with 
\begin{equation}
  \label{mGc}
m_{G}^{2}\ =\ (\lambda +\sigma  \xi g^{2})v^{2}\: +\: m^{2}\; , \qquad
m_{c}^{2}\ =\ \xi g^{2}v^{2}\; .
\end{equation} 
In the limit $\sigma \to 1$, it is $\bar{m}_{G}^{2} \to m_{G}^{2}$ and
$\bar{m}_{c}^{2}  \to  m_{c}^{2}$. Moreover,  in  the  same limit,  we
recover the  more familiar  expressions of the  $R_\xi$ gauge  for the
propagators:
\begin{equation}\label{eqn:AGpropsigmaone}
D_{G}(k)\ =\ \frac{1}{k^{2}-m_{G}^{2}}, \qquad 
\Delta_{\mu \nu}(k)\ =\ \frac{1}{k^{2}-m_{A}^{2}}
\left(-g_{\mu \nu}\: +\: (1-
\xi)\frac{k_{\mu}k_{\nu}}{k^{2}-m_{c}^{2}}\right)\, ,   
\end{equation}
whereas the ghost propagator
\begin{equation}
D_{c}(k)\ =\ \frac{1}{k^{2}-m_{c}^{2}}
\end{equation}
does not depend on the GF parameter $\sigma$; it only depends on $\xi$
through $m^2_c$. Finally, the  fermion and Higgs-boson propagators are
gauge independent, given by
\begin{equation}
  \label{eqn:fermhiggsghostprops}  
S_{1,2}(k)\ =\ \frac{1}{k_{\mu}\gamma^{\mu}-m_{f_{1,2}}}\ , \qquad 
D_{H}(k)\ =\ \frac{1}{k^{2}-m_{H}^{2}}\ , 
\end{equation}  
where
\begin{equation}\label{eqn:masses}
m_{f_{1,2}}\ =\ h_{1,2}v\;,\qquad
m_{H}^{2}\ =\ 3\lambda v^{2}\: +\: m^{2}
\end{equation} 
are the fermion and Higgs-boson masses, respectively.

\setcounter{equation}{0}
\section{The 1-Loop Effective Potential in the {\boldmath $R_{\xi / \sigma}$}
  Gauge}\label{EP}

In  order to  calculate the  1-loop  effective potential,  we use  the
functional expression~\cite{Jackiw,Zinn-Justin}:
\begin{equation}
  \label{V1loop}
V^{\mathrm{1-loop}}_{\mathrm{eff}}\ =\ 
-\,C_s\,\frac{i \hbar}{2}\ \Big( {\rm Tr}\, \ln S^{(2)}(v)\: -\: 
{\rm Tr}\, \ln S^{(2)}(0)\, \Big)\; ,
\end{equation}
where $S^{(2)}$  is the second derivative  of the classical  action $S =
\int d^4 x\; {\cal L}$, i.e.
\begin{equation}
S^{(2)}(v)\  =\   \frac{\delta^2\,S}{\delta  \phi   (x_1)\delta  \phi
(x_{2})}\;\bigg|_{\Phi = v/\sqrt{2}}\ .
\end{equation}
In the above,  $\phi$ collectively denotes each of  the quantum fields
$$\Big\{ G,\, H,\,  A^{\mu},\, c,\, \bar{c},\, f_{1,2L},\, f_{1,2R},\,
\bar{f}_{1,2L},\, \bar{f}_{1,2R} \Big\}$$ and  $C_s = +1~(-1)$ for one
real degree  of freedom  that obeys the  Bose--Einstein (Fermi--Dirac)
statistics.   Moreover,  the  trace  ${\rm Tr}$  in~(\ref{V1loop})  is
understood to act over all space and internal degrees of freedom.  For
our purposes, a more convenient representation of~(\ref{V1loop}) is
\begin{equation}
  \label{V1loopeff}
V^{\mathrm{1-loop}}_{\mathrm{eff}}\ =\
-\; C_s\ \frac{i}{2}\, 
\int^{1}_{0} dx\; {\rm Tr}\,\bigg[\;\frac{S^{(2)}(v)\: -\: 
S^{(2)}(0)}{x \left(
  S^{(2)}(v)\: -\:  S^{(2)}(0) \right)\: +\: S^{(2)}(0)}\; \bigg]\; .
\end{equation} 
In  momentum space  of $n  = 4  - 2\varepsilon$  dimension,  this last
expression becomes
\begin{equation}
  \label{V1loopk}
V^{\mathrm{1-loop}}_{\mathrm{eff}}\ =\
-\; C_s\ \frac{i}{2}\, 
\int^{1}_{0} dx\ \int \frac{d^n k}{(2\pi)^n}\
{\rm tr}\,\bigg[\; \frac{S^{(2)}(v)\: -\:  S^{(2)}(0)}{x \left(
  S^{(2)}(v)\: -\:  S^{(2)}(0) \right)\: +\: S^{(2)}(0)}\; \bigg]\; 
\end{equation} 
and ${\rm tr}$ now symbolizes the trace that should be taken only over
internal  degrees of  freedom, e.g.~over  polarizations for  the gauge
fields and over spinor components for the fermions.

Applying  now~(\ref{V1loopk}), we may  calculate the  1-loop effective
potential  by appropriately  taking  into account  all quantum  fields
$\phi$ present in the theory.   In this way, we find using dimensional
regularization        (DR)       that       for        $\sigma       =
1$~\cite{Kastening,Binosi}\footnote{Here  we seize the  opportunity to
correct several typos that occurred in Eq.~(A3) of~\cite{Binosi}},
\begin{eqnarray}
  \label{eqn:Vsigma1}
V^{\mathrm{1-loop}}_{\mathrm{eff}(\sigma=1)} \!& = &\! \frac{1}{64\pi^{2}}
 \Bigg[\, m_{H}^{4} \left( \ln\frac{m_{H}^{2}}{\bar{\mu}^{2}}
 -\frac{3}{2}\right) + m_{G}^{4}
 \left(\ln\frac{m_{G}^{2}}{\bar{\mu}^{2}} - \frac{3}{2}\right) +
 m_{A}^{4} \left( 3\ln\frac{m_{A}^{2}}{\bar{\mu}^{2}} -
 \frac{5}{2}\right) \nonumber\\
 && - m_{c}^{4} \left( \ln\frac{m_{c}^{2}}{\bar{\mu}^{2}} 
 - \frac{3}{2}\right) - 4\sum_{i=1,2}
 m_{f_{i}}^{4}\left( \ln\frac{m_{f_{i}}^{2}}{\bar{\mu}^{2}} -1 \right)
 - m^{4} \left( 2\ln\frac{m^{2}}{\bar{\mu}^{2}} - 3 \right)
 \nonumber\\ 
&& -\frac{1}{\varepsilon} \left(m_{H}^{4} + m_{G}^{4} +
 3m_{A}^{4} - m_{c}^{4} - 4\sum_{i=1}^2 m_{f_{i}}^{4} - 2m^{4}\right)\,
 \Bigg]\; ,
\end{eqnarray}
where $\ln\bar{\mu}=-\gamma +\ln 4\pi\mu^{2}$, $\gamma \approx 0.5772$
is the Euler-Mascheroni constant  and $\mu$ the renormalization scale.
Instead, for  $\sigma \neq  1$, the 1-loop  effective potential  in DR
reads:
\begin{eqnarray} 
  \label{eqn:Vsigmanot1} 
V^{\mathrm{1-loop}}_{\mathrm{eff}(\sigma \neq 1)} \!&=&\!
\frac{1}{64\pi^{2}} \Bigg[\, m_{H}^{4} \left(
\ln\frac{m_{H}^{2}}{\bar{\mu}^{2}} -\frac{3}{2}\right) +
\bar{m}_{G}^{4} \left(\ln\frac{\bar{m}_{G}^{2}}{\bar{\mu}^{2}}
-\frac{3}{2}\right) \nonumber\\ 
& & + m_{A}^{4} \left( 3\ln\frac{m_{A}^{2}}{\bar{\mu}^{2}} -
\frac{5}{2}\right) + \bar{m}_{c}^{4}
\left(\ln\frac{\bar{m}_{c}^{2}}{\bar{\mu}^{2}} - \frac{3}{2}\right) -
2m_{c}^{4} \left( \ln\frac{m_{c}^{2}}{\bar{\mu}^{2}} -
\frac{3}{2}\right) \nonumber\\ 
& & - 4\sum_{i=1}^2 m_{f_{i}}^{4}\left(
\ln\frac{m_{f_{i}}^{2}}{\bar{\mu}^{2}} -1 \right) - m^{4} \left(
2\ln\frac{m^{2}}{\bar{\mu}^{2}} - 3 \right) \nonumber\\ 
& & -\frac{1}{\varepsilon} \left(m_{H}^{4} + \bar{m}_{G}^{4} +
3m_{A}^{4} + \bar{m}_{c}^{4} - 2m_{c}^{4}  - 4\sum_{i=1}^2 m_{f_{i}}^{4} -
2m^{4}\right)\, \Bigg]\; .  
\end{eqnarray} 
We     note     that    the     limit     $\sigma     \to    1$     of
$V^{\mathrm{1-loop}}_{\mathrm{eff}(\sigma  \neq  1)}$  does equal  the
result  of  $V^{\mathrm{1-loop}}_{\mathrm{eff}(\sigma=  1)}$ given  in
(\ref{eqn:Vsigma1}). To  the best of our knowledge,  the result stated
in~(\ref{eqn:Vsigmanot1})  has   not  been  reported   before  in  the
literature, and remains central for our discussion in the next section
concerning  the gauge  dependence of  the effective  potential  in the
$R_{\xi / \sigma}$ gauge.

\setcounter{equation}{0}
\section{The Gauge Independence of the Vacuum Energy}\label{GFP}

In this section, we study the gauge dependence of the 1-loop effective
potential $V^{\mathrm{1-loop}}_{\mathrm{eff}(\sigma \neq 1)}$ given in
(\ref{eqn:Vsigmanot1}).  A  useful theoretical tool  for investigating
this is the  Nielsen identity which we apply here for  the case of the
non-linear $R_{\xi/\sigma}$  gauges.  In  particular, we show  how the
vacuum energy is a gauge-independent quantity, as required by the NI.

\subsection{The Nielsen Identity for the {\boldmath $R_{\xi /
      \sigma}$} Gauge}

Since the $R_{\xi/\sigma}$ gauge involves two GF parameters, $\xi$ and
$\sigma$,  the  original version  of  the  NI~\cite{Nielsen} needs  be
appropriately  extended.   First,  we  note  that  the  Abelian  Higgs
Lagrangian (\ref{eqn:L}) is invariant under the BRS transformations of
(\ref{eqn:BRStransformations}).   To derive the  NI, we  follow Piguet
and Sibold~\cite{PS} and promote the two GFPs, $\xi$ and $\sigma^{-1}$
(we  use  $\sigma^{-1}$  rather  than  $\sigma$  for  simplicity),  to
non-propagating  fields  attributing   their  own  anti-commuting  BRS
sources, $\eta_{\xi}$ and $\eta_{\sigma}$,~i.e.
\begin{eqnarray}
s \xi\ =\ \eta_{\xi}\;, & \qquad s \eta_{\xi}\ =\ 0\; , &  \nonumber\\
s \sigma^{-1}\ =\ \eta_{\sigma}\;, & \qquad s \eta_{\sigma}\ =\ 0\; .& 
\qquad 
\end{eqnarray}
These equations  combined with the BRS transformations  form the basis
of   the   extended    BRS   (eBRS)   transformations.    Under   eBRS
transformations,  $\mathcal{L}_{\rm inv}$ and  $\mathcal{L}_{\rm FP}$,
given in (\ref{eqn:L_{I}}) and (\ref{eqn:L_{FP}}) respectively, remain
invariant,  but not  $\mathcal{L}_{\rm  GF}$ [cf.~(\ref{eqn:L_{GF}})].
This  can be  cured by  adding  an extra  term which  is essential  to
maintain eBRS invariance, i.e.
\begin{eqnarray}
\mathcal{L}_{\rm N}\ =\ \frac{1}{2}\left(\eta_{\xi} \sigma^{-1} + \xi
\eta_{\sigma}\right) \bar{c} B  
\end{eqnarray}
The complete Lagrangian of the model has now been extended as follows:
\begin{eqnarray}
  \label{Lfull}
\mathcal{L}\! &=&\! \mathcal{L}_{\rm inv}\: +\: \mathcal{L}_{\rm GF}\: +\:
\mathcal{L}_{\rm FP}\: +\: \mathcal{L}_{\rm N}\nonumber\\
\! &=&\! \mathcal{L}_{\rm inv}\: +\:
\frac{\xi}{2 \sigma}B^{2}\: +\: BF\: -\: \bar{c}(sF)\: +\:
\frac{1}{2}\left(\eta_{\xi} \sigma^{-1}\: +\: \xi \eta_{\sigma}\right)
\bar{c} B\; ,
\end{eqnarray}
which   is   invariant   under   eBRS  transformations.    

Given the Lagrangian~(\ref{Lfull}),  the generating functional $Z$ for
the connected Green functions is obtained by
\begin{eqnarray}
\exp (iZ) \!&=&\! \int  [dA_{\mu}] [dc] [d\bar{c}] [dH] [dG] [dB] [df_L] [df_R]
      [d\bar{f}_L] [d\bar{f}_R]\ \exp \bigg[\, i \int d^4 x\, 
\Big( \mathcal{L}\:  +\: J^{\mu}_A A_{\mu}\nonumber\\ 
\!&&\! +\:
 J_c c\: +\: J_{\bar{c}} \bar{c}\: +\: J_H H\: +\: 
 J_G G\: +\:  J_{f_L} f_L\: +\: J_{f_R} f_R\: +\:
 J_{\bar{f}_L}\bar{f}_L\: +\: J_{\bar{f}_R}\bar{f}_R\nonumber\\ 
\!&&\! +\: K_H sH\: +\: K_G sG\: +\: 
K_{\bar{c}} s\bar{c}\: +\: K_{f_L} sf_L\: +\: 
K_{\bar{f}_L} s\bar{f}_L  \Big)\, \bigg]
\end{eqnarray} 
where $J_\phi$  are the  sources of all  quantum fields $\phi$  in the
model and  $K_\phi$ are  their respective sources  coupled to  the BRS
transforms of $\phi$.  We assign no $J$-source to  the auxiliary field
$B$, since this will be identical to the source $K_{\bar{c}}$.
We may now integrate out the auxiliary field $B$ using its equation
of motion,
\begin{equation}
  \label{Bfield}
B\ =\ -\frac{\sigma}{\xi}\;\bigg[\, F\: +\: \frac{1}{2}\left(\eta_{\xi}
  \sigma^{-1}\, +\: \xi \eta_{\sigma}\right) \bar{c}\: +\: 
K_{\bar{c}}\, \bigg]\; . 
\end{equation}
Substituting (\ref{Bfield}) into 
$\mathcal{L}_{\rm GF}$ and $\mathcal{L}_{N}$ yields
\begin{equation}
\mathcal{L}_{\rm GF}\ =\ -\frac{\sigma}{2 \xi}F^{2}\: +\: \frac{\sigma}{2
\xi}K_{\bar{c}}^{2}\; , \qquad  \mathcal{L}_{N}\ =\ -\frac{\sigma}{2
\xi}\left(\eta_{\xi} \sigma^{-1}\: +\; \xi \eta_{\sigma}\right)
\bar{c} F\; .
\end{equation}
The price one has  to pay here by the elimination of  the $B$ field is
that the eBRS algebra is  no longer nilpotent for the anti-ghost field
$\bar{c}$, i.e.
\begin{equation}
  \label{scbar}
s^{2} \bar{c}\ =\ \frac{\sigma}{\xi}\; \bigg[\, \frac{\sigma}{2 \xi}
\left(\eta_{\xi} \sigma^{-1}\, +\: \xi \eta_{\sigma}\right)
\Big(F\, +\, K_{\bar{c}}\Big)\: -\; (sF)\,\bigg]\; .
\end{equation}
However, the  eBRS algebra closes on-shell, i.e.~$s^{2}  \bar{c} = 0$,
after  the  equation  of  motion  of  $\bar{c}$ is  used  on  the  RHS
of~(\ref{scbar}) while setting $J_{\bar{c}}=0$.

We  may  now derive  the  NI  by first  noticing  that  under an  eBRS
transformation  the  generating  functional  $Z$, with  the  $B$-field
appropriately eliminated, obeys the symmetry relation~\cite{Binosi}:
\begin{eqnarray}
  \label{Zomega}
&&
 Z[J^{\mu}_{A},J_{c},J_{\bar{c}},J_{H},J_{G},J_{f_L},J_{f_R},
 J_{\bar{f}_L},J_{\bar{f}_R},K_{H},K_{G},K_{\bar{c}},K_{f_L},
 K_{\bar{f}_L}; 
 \xi , \sigma^{-1}, \eta_{\xi}, \eta_{\sigma}]\ =\ \nonumber\\ 
&&
 Z[J^{\mu}_{A},J_{c}-\omega \partial_{\mu}J^{\mu}_{A}
 ,J_{\bar{c}},J_{H},J_{G},J_{f_L},J_{f_R},
 J_{\bar{f}_L},J_{\bar{f}_R},K_{H} + 
 \omega J_{H},K_{G}+ \omega J_{G}, \nonumber\\ 
&& K_{\bar{c}} + \omega J_{\bar{c}},K_{f_L} + \omega J_{f_L},
 K_{\bar{f}_L} + \omega J_{\bar{f}_L} ; \xi- \omega
 \eta_{\xi}, \sigma^{-1} - \omega \eta_{\sigma}, \eta_{\xi},
 \eta_{\sigma}]\ .
\end{eqnarray}
where  $\omega$ is  an arbitrary  anticommuting parameter  required to
conserve the FP ghost charge  $Q_{\rm FP}$, for which $Q_{\rm FP}(s) =
1$   and  $Q_{\rm   FP}(\omega)  =-1$.    We  then   expand   the  RHS
of~(\ref{Zomega})  to  order~$\omega$  to  obtain  the  Slavnov-Taylor
identity
\begin{equation} 
  \label{eqn:4.12}
-(\partial_{\mu}J^{\mu}_{A})\frac{\delta Z}{\delta J_{c}} +
 J_{H}\frac{\delta Z}{\delta K_{H}} + J_{G}\frac{\delta Z}{\delta
 K_{G}} + J_{\bar{c}}\frac{\delta Z}{\delta K_{\bar{c}}} +
 J_{f_L}\frac{\delta Z}{\delta K_{f_L}} +
 J_{\bar{f}_L}\frac{\delta Z}{\delta K_{\bar{f}_L}}-
 \eta_{\xi}\partial_{\xi} Z 
  - \eta_{\sigma}\partial_{\sigma^{-1}}Z = 0.
\end{equation}
The  above identity can  now be  rewritten in  terms of  the effective
action $\g$, which is defined via the Legendre transformation,
\begin{equation} 
  \label{eqn:effaction}
\g [\phi_{\rm cl}, K_{\phi}; \xi ,\sigma^{-1}, \eta_{\xi},
\eta_{\sigma}]\ =\ Z[J_{\phi}, K_{\phi}; \xi ,\sigma^{-1}, \eta_{\xi},
\eta_{\sigma}]\ -\ \int d^{4}x J_{\phi}\phi_{\rm cl}\; ,
\end{equation}
where $\phi_{\rm cl}$ is the classical field defined as $\phi_{\rm cl}
= \delta  Z/\delta J_{\phi}$.  In addition,  it can be  shown that the
following relations are satisfied by the effective action:
\begin{equation}
\frac{\delta \g}{\delta \phi_{\rm cl}}\ =\ -J_{\phi}\; ,\qquad
\frac{\delta \g}{\delta K_{\phi}}\ =\ \frac{\delta Z}{\delta
  K_{\phi}}\;, \qquad 
\partial _{\xi} \g\ =\ \partial_{\xi} Z, \qquad 
\partial _{\sigma^{-1}} \g\ =\ \partial_{\sigma^{-1}} Z\; .
\end{equation}  
With the help of these relations, (\ref{eqn:4.12}) becomes,
\begin{eqnarray} 
\eta_{\xi}\partial_{\xi} \g\: +\:
\eta_{\sigma}\partial_{\sigma^{-1}} \g \!&=&\! -\;\Bigg(
\frac{\delta \g}{\delta
A^{\mu}_{\rm cl}}\partial_{\mu}c_{\rm cl}\ +\ \frac{\delta \g}{\delta
  \bar{c}_{\rm cl}}
\frac{\delta \g}{\delta
K_{\bar{c}}}\ +\ \frac{\delta \g}{\delta
{H}_{\rm cl}}\frac{\delta \g}{\delta K_{H}}\ +\ \frac{\delta
\g}{\delta G_{\rm cl}}\frac{\delta \g}{\delta K_{G}} \nonumber\\ 
& & +\ \frac{\delta \g}{\delta f_{{\rm cl}L}}
\frac{\delta \g}{\delta
K_{f_L}}\ +\ \frac{\delta \g}{\delta
\bar{f}_{{\rm cl}L}}\frac{\delta \g}{\delta
K_{\bar{f}_L}}\Bigg)\; .
\end{eqnarray}
Differentiating  with respect to  $\eta_{\xi}$ or  $\eta_{\sigma}$ and
then setting $\eta_{\xi}  = \eta_{\sigma}=0$ gives rise to  the NI for
the Abelian Higgs  model under study:\footnote{The present derivation
for the  $R_{\xi/\sigma}$ gauge  extends the previous  result obtained
in~\cite{HK}.}
\begin{eqnarray}
  \label{eqn:NI}
\partial_{x} \g |_{\eta_{\xi}= \eta_{\sigma}=0}  \!&=&\!
-\; \partial_{\eta_{x}}\Bigg( \frac{\delta \g}{\delta
A^{\mu}_{\rm cl}}\partial_{\mu}c_{\rm cl}\ +\ \frac{\delta \g}{\delta
  \bar{c}_{\rm cl}}
\frac{\delta \g}{\delta K_{\bar{c}}}\ +\ 
\frac{\delta \g}{\delta H_{\rm cl}}\frac{\delta \g}{\delta K_H}\ +\ 
\frac{\delta \g }{\delta G_{\rm cl}}\frac{\delta \g}{\delta K_{G}}\nonumber\\ 
&& +\ \frac{\delta \g}{\delta f_{{\rm cl}L}}\frac{\delta \g}{\delta
K_{f_L}}\ +\ \frac{\delta \g}{\delta \bar{f}_{{\rm cl}L}}
\frac{\delta \g}{\delta K_{\bar{f}_L}} \Bigg)
\Bigg|_{\eta_{\xi}= \eta_{\sigma}=0}\; ,
\end{eqnarray}
where  $x =  \xi$ or  $\sigma^{-1}$. An  immediate consequence  of the
NI~(\ref{eqn:NI}) is  that the effective action is  independent on the
GF parameters $\xi$ and  $\sigma$, once the extremal condition $\delta
\g /\delta \phi_{\rm  cl} = 0$ is satisfied for  each of the classical
fields $\phi_{\rm cl}$ of the theory.

\subsection{Gauge Dependence of the 1-loop Effective Potential}

\begin{figure}
\qquad \includegraphics{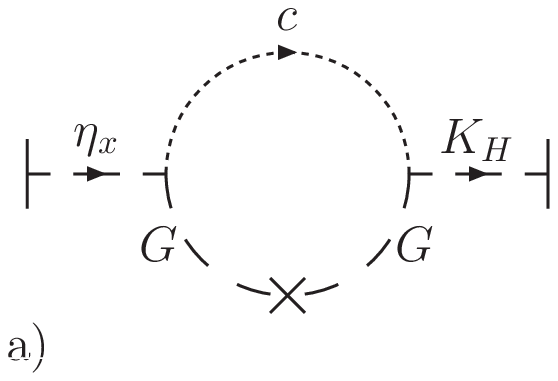} \qquad \qquad \includegraphics{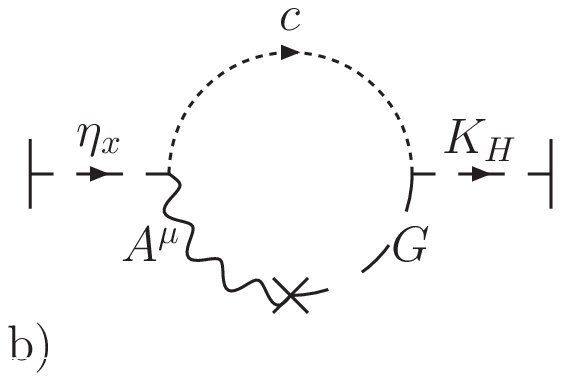} 
\caption{\it  The  Feynman  diagrams  which  contribute  to  the  term
$\partial_{\eta_{x}}(\delta     \g^{\mathrm{1-loop}}/\delta    K_{H})$
in~(\ref{etaKH}). The second graph is required for $\sigma \neq 1$ and
takes account of the gauge dependence due to the Goldstone--gauge-boson
mixing.}\label{fig:etaKH}
\end{figure}

The  NI~(\ref{eqn:NI}) also  holds  true for  the effective  potential
$V_{\rm eff}$.  For translationally  invariant solutions of $\delta \g
/\delta  \phi_{\rm cl}  = 0$,  the  effective potential  has a  simple
connection  with  the  effective  action:  $V_{\rm eff}  =  -  d^4  \g
/dx^4$. Since only the Higgs field $H_{\rm cl}$ can provide a non-zero
translational invariant solution to $\delta\g /\delta H_{\rm cl} = 0$,
the NI for the 1-loop effective potential simplifies considerably to
\begin{equation}
  \label{etaKH}
\partial_{x}   V_{\rm   eff}^{\mathrm{1-loop}}   \   =\   \frac{\delta
V^{\mathrm{tree}}}{\delta                                    {H}_{\rm cl}}\
\partial_{\eta_{x}}\left(\frac{\delta      \g^{\mathrm{1-loop}}}{\delta
K_{H}}\right)\; ,
\end{equation}
where $x =  \xi$ or  $\sigma^{-1}$, and
\begin{equation}
\frac{\delta V^{\mathrm{tree}}}{\delta {H}_{\rm cl}}\ =\
-v(m^{2}+\lambda v^{2})\; .
\end{equation} 
The    Feynman    graphs    that     give    rise    to    the    term
$\partial_{\eta_{x}}(\delta     \g^{\mathrm{1-loop}}/\delta    K_{H})$
in~(\ref{etaKH}) are shown  in Fig.~\ref{fig:etaKH}.  Observe that for
$\sigma = 1$, only the graph in Fig.~\ref{fig:etaKH}(a) contributes in
this case.

%%%%%%%%%%%%%%%%%%%%%%%%%%%%%%%%%%%%%%%%%%%%%%%%%%%%%%%%%%%%%
%%%%%%%%%%%%%%%%%%%%%%%%%%%%%%%%%%%%%%%%%%%%%%%%%%%%%%%%%%%%

Let us  first consider  the $R_\xi$ set  of gauges,  where $\sigma=1$.
Calculating  the  Feynman  graph  in Fig.~\ref{fig:etaKH}(a)  at  zero
external momentum, we obtain
\begin{equation}
\partial_{\eta_{\xi}}\left(\frac{\delta 
\g^{\mathrm{1-loop}}}{\delta K_{H}} \right)\ =\ \frac{g^{2}v}{32
\pi^{2}}\, \bigg[\, \frac{1}{\lambda v^{2} + m^{2}}\left(m_{c}^{2}\, \ln
\frac{m_{c}^{2}}{\bar{\mu}^{2}}\: -\: m_{G}^{2}\, \ln
\frac{m_{G}^{2}}{\bar{\mu}^{2}} \right)\ +\: 1\: +\ \frac{1}{\varepsilon}\,
\bigg]\; .
\end{equation}
Consequently, the gauge dependence of the 1-loop effective potential
is determined by means of the NI~(\ref{etaKH}):
\begin{eqnarray} 
\label{NI1} 
\partial_{\xi} \Big(V^{\mathrm{1-loop}}_{\rm eff}\Big)_{R_\xi} 
\!& = &\! \frac{g^{2}v^{2}}{32 \pi^{2}}\, \bigg[\, \xi
 g^{2} v^{2} \bigg(\ln \frac{\lambda v^{2} + m^{2} + \xi g^{2}
 v^{2}}{\bar{\mu}^{2}}\ -\ \ln \frac{\xi g^{2}
 v^{2}}{\bar{\mu}^{2}}\,\bigg) \\ 
\!&&\! +\ (\lambda v^{2} +
 m^{2})\,\bigg(\ln \frac{\lambda v^{2} + m^{2} + \xi g^{2}
 v^{2}}{\bar{\mu}^{2}}\ -\: 1\: -\ \frac{1}{\varepsilon}\, \bigg)\,
 \bigg]\; .
 \nonumber
\end{eqnarray}
It is easy to check that the above result coincides with that obtained
after  differentiating  the analytical  expression  for the  effective
potential in~ (\ref{eqn:Vsigma1}) with respect to $\xi$.

We now turn to the non-trivial case of $\sigma \neq 1$.  In this case,
one should  consider the mixing  between the would-be  Goldstone boson
$G$ and the gauge  boson $A^\mu$, as shown in Fig.~\ref{fig:etaKH}(b).
In  particular,  taking   both  graphs  of  Fig.~\ref{fig:etaKH}  into
account,  we  may calculate  the  gauge  dependence  of the  effective
potential on the two GF parameters $\xi$ and $\sigma$:
\begin{eqnarray}        
  \label{NIxi}
\partial_{\xi} \Big(V^{\mathrm{1-loop}}_{\rm eff}\Big)_{R_{\xi/\sigma}} 
\!&=&\! \frac{g^{2}v^{2}}{32 \pi^{2}}
\bigg\{\; \xi g^{2} v^{2}\; \bigg( \ln \frac{\lambda v^{2} + m^{2} +
\sigma \xi g^{2} v^{2}}{\sigma \bar{\mu}^{2}}\ -\ \ln \frac{ \xi g^{2}
v^{2}}{\bar{\mu}^{2}}\, \bigg) \nonumber\\ 
\!&&\!\hspace{-6mm} +\ \frac{1}{2}\; (\lambda
v^{2} + m^{2})^{1/2}\; \bigg[\; \lambda v^{2}\: +\: m^{2}\: +\: 
\frac{4\xi}{\sigma}\; g^{2} v^{2}(\sigma -1) \bigg]^{1/2}\; \ln
\frac{\bar{m}_{G}^{2}}{\bar{m}_{c}^{2}} \nonumber\\ 
\!&&\!\hspace{-6mm} +\ (\lambda v^{2} + m^{2})\;
\bigg(\,\ln \frac{\bar{m}_{G}^{2}}{\bar{\mu}^{2}}\ +\ \ln
\frac{\bar{m}_{c}^{2}}{\bar{\mu}^{2}}\ +\ \frac{1}{2}\ln
\frac{{m}_{G}^{2}}{\bar{\mu}^{2}}\ +\ \frac{1}{2}
\ln \frac{{m}_{c}^{2}}{\bar{\mu}^{2}}\: -\ 2\ 
+\ \frac{1}{\sigma}\,\bigg)\nonumber\\ 
\!&&\!\hspace{-6mm} -\ \frac{1}{\varepsilon}\;
(\lambda v^{2} + m^{2})\bigg(2-\frac{1}{\sigma}\bigg)\; \bigg\}\;,\\[3mm]
  \label{NIsigma}
\partial_{\sigma^{-1}} \Big(V^{\mathrm{1-loop}}_{\rm
 eff}\Big)_{R_{\xi/\sigma}} 
 \!&=&\! \frac{\xi g^{2}
v^{2}}{32 \pi^{2}}\ (\lambda v^{2} + m^{2})\; \bigg[\, 
\frac{1}{\bar{m}_{G}^{2} - \bar{m}_{c}^{2}}\ \bigg(\,
\bar{m}_{G}^{2}\; \ln \frac{\bar{m}_{G}^{2}}{\bar{\mu}^{2}}\ 
-\ \bar{m}_{c}^{2}\; \ln \frac{\bar{m}_{c}^{2}}{\bar{\mu}^{2}}\, 
\bigg) \nonumber\\ 
\!&&\!\hspace{-6mm} -\ 1\: -\ \frac{1}{\varepsilon}\; \bigg]\; . 
\end{eqnarray}
Again,  we have  checked  that the  same  result is  obtained, if  the
analytical   expression    for   the   1-loop    effective   potential
in~(\ref{eqn:Vsigmanot1})  is differentiated  with respect  to  the GF
parameters $\xi$ and $\sigma^{-1}$, in agreement with the NI.

From~(\ref{NIxi})  and~(\ref{NIsigma}), it  is then  not  difficult to
prove the gauge independence of  the 1-loop effective potential at its
extrema. To zeroth  order of loop expansion, the  extrema of the Higgs
potential, determined from (\ref{eqn:L_{1}}), are
\begin{equation} 
v\ =\ 0\;, \qquad v^{2}\ =\ - \frac{m^{2}}{\lambda}\ .
\end{equation}
Up  to  an  unspecified  additive  cosmological  constant,  the  first
solution corresponding to a  local maximum gives a vanishing effective
potential,   i.e.~$V^{\mathrm{1-loop}}_{\mathrm{eff}}   (v=0)  =   0$,
whereas the second one,  $v^{2}\ =\ - m^{2}/\lambda$ which corresponds
to the  global minimum of  the effective potential to  ${\cal O}(\hbar
)$, gives
\begin{eqnarray}
\Big(V^{\mathrm{1-loop}}_{\mathrm{eff}} \Big)_{R_{\xi/\sigma}} \!&=&\!
\frac{m^{4}}{64\pi^{2}}\ \bigg[\, 4\ln\frac{-2m^{2}}{\bar{\mu}^{2}}\ -\
2\ln\frac{m^{2}}{\bar{\mu}^{2}}\ -\ 3\ +\
\frac{g^{4}}{\lambda^{2}}\;
\bigg(\,3\ln\frac{g^{2}(-m^{2}/\lambda)}{\bar{\mu}^{2}}\ 
-\ \frac{5}{2}\,\bigg) \nonumber\\ 
\!&&\!\hspace{-15mm} -\ 4\;\sum_{i=1,2}\;
\frac{h_{i}^{4}}{\lambda^{2}}\
\bigg(\,\ln\frac{h_{i}^{2}(-m^{2}/\lambda)}{\bar{\mu}^{2}} - 1
\bigg)\ -\ \frac{1}{\varepsilon}\bigg(\,2\: +\
\frac{3g^{2}}{\lambda^{2}}\ - 4\,
\frac{\sum_{i=1,2}\; h_{i}^{4}}{\lambda^{2}}\, \bigg)\; \bigg]\; .\qquad
\end{eqnarray}
This completes our  proof that, up to 1-loop  level, the vacuum energy
of  the  Abelian  Higgs  model  is gauge  invariant  in  this  general
$R_{\xi/\sigma}$ class of non-linear gauges.

\setcounter{equation}{0}
\section{1-loop Renormalization}\label{1LR}

In  this section,  we  present the  1-loop  constant and  wavefunction
renormalizations  in  the non-linear  $R_{\xi/\sigma}$  gauges in  the
$\overline{\rm  MS}$  scheme.   To  this  end, we  use  the  so-called
displacement operator formalism, or  $D$-formalism in short, which was
developed   in~\cite{Binosi}    as   an   alternative    approach   to
systematically   performing   renormalization   to   all   orders   in
perturbation theory.

According  to  the   $D$-formalism,  the  renormalized  1PI  $n$-point
correlation  functions,  denoted  hereafter  with a  script  $R$,  are
related to the unrenormalized ones through:
\begin{equation}
  \label{Dformalism}
\phi_{R}^{n}\, \g^{R}_{\phi^{n}}(\lambda_{R}, \sigma_{R}, \xi_{R};
\mu)\ =\ e^{D}\,\Big( \phi_{R}^{n} \g_{\phi^{n}}(\lambda_{R},
\sigma_{R}, \xi_{R}; \mu, \varepsilon) \Big)\; ,
\end{equation}
where $D$ is the displacement operator that takes on the form,
\begin{equation}
D\ =\ \delta \phi \frac{ \partial}{ \partial \phi_{R}}\: +\: \delta
\lambda \frac{ \partial}{ \partial \lambda_{R}}\: +\: \delta \sigma \frac{
\partial}{ \partial \sigma_{R}}\: +\: \delta \xi \frac{ \partial}{ \partial
\xi_{R}}\; ,
\end{equation}
where $\phi$ represents all the  fields in the model and $\lambda$ all
the  coupling   and  mass  parameters,   i.e.~$\lambda,\  m^{2},\  g,\
h_{1,2}$, including the  VEV $v$ of the Higgs  field. In addition, the
counterterm  renormalizations, $\delta\phi$, $\delta\lambda$  etc, are
defined as, $\delta \phi = \phi - \phi_R = (Z^{1/2}_\phi - 1) \phi_R$,
$\delta\lambda = \lambda -\lambda_R  = (Z_\lambda - 1) \lambda_R$ etc.
Since the $R_{\xi/\sigma}$ gauge obeys  the HLET, $v$ does not need to
have  its  own  counterterm  and  it is  renormalized  via  the  Higgs
wavefunction, i.e.~$v = Z_{H}^{\frac{1}{2}}v_{R}$.

We  may  now  perform  a  loopwise expansion  of  the  operator  $e^D$
in~(\ref{Dformalism}),
\begin{equation}
e^{D} = 1 + D^{(1)} + ( D^{(2)} + \frac{1}{2}D^{(1)2}) + ...
\end{equation}
where the superscript $(n)$ on $D$ denotes the loop order, i.e.
\begin{equation}
D^{(n)}\ =\ \delta \phi^{(n)} \frac{ \partial}{ \partial \phi_{R}}\: +\:
\delta \lambda^{(n)} \frac{ \partial}{ \partial \lambda_{R}}\: +\: \delta
\sigma^{(n)} \frac{ \partial}{ \partial \sigma_{R}}\: +\: \delta \xi^{(n)}
\frac{ \partial}{ \partial \xi_{R}}\; .
\end{equation}
Correspondingly,   the   parameter   or  counterterm   shifts   $\delta
\phi^{(n)},\  \delta \lambda^{(n)},\ \delta  \sigma^{(n)}$  and $\delta
\xi^{(n)}$ are loopwise defined as
\begin{equation}
\delta \phi^{(n)}\ =\ Z_{\phi}^{\frac{1}{2}(n)}\phi_{R}, \qquad \delta
\lambda^{(n)}\ =\ Z_{\lambda}^{(n)}\lambda_{R}, \qquad \delta
\sigma^{(n)}\ =\ Z_{\sigma}^{(n)}\sigma_{R}, \qquad \delta \xi^{(n)}\ =\
Z_{\xi}^{(n)}\xi_{R}\ .
\end{equation}
Applying the $D$-formalism to 1-loop, we have
\begin{equation}
  \label{D1loop}
\phi_{R}^{n} \Gamma^{R(1)}_{\phi^{n}}(\lambda_{R}, \sigma_{R},
\xi_{R}; \mu)\ =\ D^{(1)}\left( \phi_{R}^{n}
\Gamma^{(0)}_{\phi^{n}}(\lambda_{R}, \sigma_{R}, \xi_{R}; \mu) \right)
+ \phi_{R}^{n} \Gamma^{(1)}_{\phi^{n}}(\lambda_{R}, \sigma_{R},
\xi_{R}; \mu, \varepsilon)\; .
\end{equation} 

Employing~(\ref{D1loop})  in the  $\overline{\rm MS}$  scheme,  we may
calculate the 1-loop constant and wavefunction renormalizations in the
non-linear $R_{\xi/\sigma}$ gauge.  These are given by
\begin{eqnarray}
  \label{Zetas}
Z_{H} & = & 1 + \frac{1}{(4\pi)^{2} \varepsilon} \left(3g^{2} + \xi g^{2}
- 2 \sum_{i=1}^2 h_{i}^{2} + \frac{\xi g^{2}}{\sigma}(\sigma-1)\right)\; ,
\nonumber\\ 
Z_{G} & = & 1 + \frac{1}{(4\pi)^{2} \varepsilon} \left(3g^{2} - 3\xi
g^{2} - 2 \sum_{i=1}^2 
h_{i}^{2} + \frac{\xi g^{2}}{\sigma}(\sigma-1)\right)\; ,
\nonumber\\ 
Z_{\sigma} & = & 1 +  \frac{1}{(4\pi)^{2} \varepsilon} \left(-3g^{2} +
3\xi g^{2} + 2\lambda + 2 \sum_{i=1}^2 h_{i}^{2} - \frac{\xi
  g^{2}}{\sigma}(\sigma-1) + \frac{\xi g^{2}}{\sigma}(\sigma-1)^{2}
\right)\; , \nonumber\\ 
Z_{\xi} & = & 1 + \frac{1}{(4\pi)^{2} \varepsilon}
\left(-\frac{14}{3}g^{2} + 2\xi g^{2} + 2\lambda + 2 \sum_{i=1}^2 h_{i}^{2} -
2\frac{\xi g^{2}}{\sigma}(\sigma-1)\right)\; , \nonumber\\ 
Z_{A} & = & 1 + \frac{1}{(4\pi)^{2} \varepsilon}
\left(-\frac{5}{3}g^{2}\right)\; , \nonumber\\ 
Z_{c} & = & 1\; , \nonumber\\
Z_{g} & = & 1 + \frac{1}{(4\pi)^{2} \varepsilon}
\left(\frac{5}{6}g^{2}\right)\; , \nonumber\\ 
Z_{\lambda} & = & 1 + \frac{1}{(4\pi)^{2} \varepsilon} \left(-6g^{2} + 10
\lambda + 4 \sum_{i=1}^2 h_{i}^{2} + \frac{3g^{2}}{\lambda} - \frac{ 4 
\sum_{i=1}^2  h_{i}^{4}}{\lambda}\right)\; , \nonumber\\ 
Z_{m^{2}} & = & 1 + \frac{1}{(4\pi)^{2} \varepsilon} \left(-3g^{2} + 4
\lambda + 2 \sum_{i=1}^2 h_{i}^{2} \right)\; , \nonumber\\ 
Z_{1,2}^{L} & = & 1 +  \frac{1}{(4\pi)^{2} \varepsilon} \left(- g^{2}\xi -
h_{1,2}^{2} + \frac{\xi g^{2}}{\sigma}(\sigma-1) \right)\; , \nonumber\\ 
Z_{1,2}^{R} & = & 1 +  \frac{1}{(4\pi)^{2} \varepsilon} \left(-
h_{1,2}^{2}\right)\; , \nonumber\\ 
Z_{h_{1,2}} & = & 1 +  \frac{1}{(4\pi)^{2} \varepsilon} \left(- \frac{3}{2}
g^{2} +  h_{1,2}^{2} +  \sum_{i=1}^{2} h_{i}^{2} + 2\frac{\xi
  g^{2}}{\sigma}(\sigma-1)\right)\; .  
\end{eqnarray} 
Observe that Green's functions in the $R_\xi$ gauge, where $\sigma_R =
1$,  would require  an additional  renormalization constant,  given by
$\delta \sigma^{(1)}  = Z_{\sigma =  1}^{(1)} - 1$, beyond  the 1-loop
level,  in order to  render them  ultra-violet finite.   For instance,
this will  be the  case, when one  is computing  $n$-point correlation
functions  that involve Higgs  bosons or  left-handed fermions  in the
Abelian Higgs  model under study.  The  only exception to  this is the
Landau gauge  $\xi =  0^+$, which does  not require  a renormalization
counterterm.  The reason  is that  unlike $\sigma$,  the  GF parameter
$\xi$ renormalizes  multiplicatively, so the condition $\xi  = 0$ will
not be  affected by renormalization.   Moreover, in the  Landau gauge,
the  role  of  the  GF  parameter $\sigma$  is  redundant,  since  any
$\sigma$-dependence of the counterterms  contains a factor of $\xi$ as
well  and   therefore  it  vanishes   in  the  limit  $\xi   \to  0^+$
[cf.~(\ref{Zetas})].

\setcounter{equation}{0}
\section{Conclusions}\label{conclusions}

We have calculated the 1-loop  effective potential of an Abelian Higgs
model  within the  $R_{\xi/\sigma}$ class  of non-linear  gauges.  The
$R_{\xi/\sigma}$ class  involves two GF parameters  $\xi$ and $\sigma$
and constitutes  a renormalizable  extension of the  Feynman--'t Hooft
$R_\xi$ set of gauges.  In  particular, it enables one to consistently
study the  gauge dependence of $n$-point  correlation functions beyond
the 1-loop level.  Another advantage of the $R_{\xi/\sigma}$ GF scheme
is  that it preserves  the Higgs-boson  low-energy theorem,  which has
found     to     have     several     applications     in     particle
phenomenology~\cite{HLET}     including     those     in     $B$-meson
physics~\cite{Bmeson}.

In the $R_{\xi/\sigma}$ class of  gauges, one has to properly consider
Goldstone--gauge-boson  mixing  effects.   Taking these  effects  into
account, we  have shown that the 1-loop  effective potential evaluated
at  its  extrema  is  independent  of  both  the  GF  parameters~$\xi$
and~$\sigma$.  This result is  in agreement with the Nielsen identity,
which  we  have  derived here  for  the  Abelian  Higgs model  in  the
$R_{\xi/\sigma}$ gauge.

It  is  important to  stress  that  the Goldstone--gauge-boson  mixing
vanishes at  the tree  level in the  Feynman--'t Hooft $R_\xi$  set of
gauges, when $\sigma = 1$, and in the Landau gauge $\xi = 0^+$. Unlike
in the Landau  gauge, however, the mixing of the  gauge field with its
associated would-be  Goldstone boson  will reappear beyond  the 1-loop
level within the frequently used  Feynman $\sigma = 1$ gauge.  We have
therefore calculated the 1-loop counterterms for couplings, masses and
field  wavefunctions  within  the  general $R_{\xi/\sigma}$  class  of
gauges, for arbitrary values of  the $\sigma$ parameter.  We hope that
the pivotal  study presented in this  paper will be  useful for future
higher-order  calculations and cross-checks  of gauge  independence of
physical  observables,  within   more  realistic  models  of  particle
physics.

\subsection*{Acknowledgements}
This work is  supported in part  by the STFC research grant: PP/D000157/1.

\end{document}